\documentclass{aa}  
\usepackage{graphicx}
\usepackage{txfonts}
\usepackage{caption}
\newcommand{\pccm}{pc\,cm$^{-3}$}
\newcommand{\h}{^{\rm h}}
\newcommand{\m}{^{\rm m}}
\newcommand{\s}{^{\rm s}}

\begin{document}

   \title{A MeerKAT view of the pulsars in the globular cluster NGC 6522}

   \author{F. Abbate
          \inst{1,2,3}
          \and
          A. Ridolfi\inst{1,2}
          \and
          P. C. C. Freire\inst{1}
          \and
          P. V. Padmanabh\inst{4,5,1}
          \and
          V. Balakrishnan\inst{1}
          \and
          S. Buchner\inst{6}
          \and
          L. Zhang\inst{7}
          \and
          M. Kramer\inst{1,8}
          \and
          B. W. Stappers\inst{8}
          \and
          E. D. Barr\inst{1}
          \and
          W. Chen\inst{1}
          \and 
          D. Champion\inst{1}
          \and
          S. Ransom\inst{9}
          \and
          A. Possenti\inst{2}
          }

   \institute{Max-Planck-Institut f\"{u}r Radioastronomie, Auf dem H\"{u}gel 69, D-53121 Bonn, Germany\label{1}
   \and
    INAF -- Osservatorio Astronomico di Cagliari, Via della Scienza 5, I-09047 Selargius (CA), Italy\label{2}
    \and
    Dipartimento di Fisica "G. Occhialini", Universit\`a degli Studi di Milano-Bicocca, Piazza della Scienza 3, I-20126 Milano, Italy\label{3}
    \and
    Max Planck Institute for Gravitational Physics (Albert Einstein Institute), D-30167 Hannover, Germany\label{4}
    \and
    Leibniz Universit\"{a}t Hannover, D-30167 Hannover, Germany\label{5}
    \and
    South African Radio Astronomy Observatory, 2 Fir Street,Cape Town, 7925, South Africa\label{6}
    \and
    National Astronomical Observatories, Chinese Academy of Sciences, A20 Datun Road, Chaoyang District, Beijing 100101, People’s Republic of China\label{7}
    \and
    Jodrell Bank Centre for Astrophysics, Department of Physics and Astronomy, The University of Manchester, Manchester M13 9PL, UK\label{8}
    \and
    National Radio Astronomy Observatory, 520 Edgemont Rd., Charlottesville, VA, 22903, USA\label{9}
    }

   \date{}

% \abstract{}{}{}{}{} 
% 5 {} token are mandatory
 
  \abstract{ 
  We present the results of observations aimed at discovering and studying pulsars in the core-collapsed globular cluster (GC) NGC 6522 performed by the MeerTIME and TRAPUM Large Survey Project with the MeerKAT telescope. We have discovered two new isolated pulsars bringing the total number of known pulsars in the cluster to six. PSR J1803$-$3002E is a mildly recycled pulsar with spin period of 17.9 ms while pulsar PSR J1803$-$3002F is a slow pulsar with spin period of 148.1 ms. The presence of isolated and slow pulsars is expected in NGC 6522 and confirms the predictions of previous theories for clusters at this stage in evolution. We further present a tentative timing solution for the millisecond pulsar (MSP) PSR J1803$-$3002C combining older observations taken with the Parkes 64m radio telescope, Murriyang. This solution implies a relatively small characteristic age of the pulsar in contrast with the old age of the GC. The presence of a slow pulsar and an apparently young MSP, both rare in GCs, suggests that their formation might be linked to the evolutionary stage of the cluster.
  }

   \maketitle
%
%-------------------------------------------------------------------

\section{Introduction}

The globular cluster (GC) NGC 6522 is located in the bulge of the Milky Way in a region with relatively low amounts of interstellar gas called Baade's Window \citep{Baade1946}. It is located at a distance of 7.3 kpc \citep{Baumgardt2021}\footnote{Adopted from the list of the structural parameters of GCs available at the website: \url{ https://people.smp.uq.edu.au/HolgerBaumgardt/globular/parameter.html}}  from the Sun and only 1 kpc from the Galactic Center. It is also one of the oldest GCs in the Milky way with an age estimate of $\sim 12.5$ Gyr \citep{Barbuy2009,Kerber2018} and it is classified as core-collapsed \citep{Trager1995}.

Like many other GCs, NGC 6522 has been observed over the years with the goal of finding pulsars. A total of 4 have been discovered so far (PSR J1803$-$3002A with the Parkes 64m ``Murriyang'' radio telescope, henceforth `Parkes telescope'', \citealt{Possenti2005}, J1803$-$3002B and C with the Green Bank telescope, \citealt{Begin_2006}) and J1803$-$3002D with MeerKAT radio telescope, \citealt{Ridolfi2021}. Of these only PSR J1803-3002A has a published timing solution \citep{Zhang2020}. 
%\citep{Possenti2005,Begin_2006,Ridolfi2021}.
These pulsars are all isolated millisecond pulsars (MSPs) with periods ranging from 4.3 ms for PSR J1803$-$3002B to 7.1 ms for PSR J1803$-$3002A. The lack of binary pulsars is quite unusual for the typical GC but not for core-collapsed clusters that seem to show a strong predominance for isolated pulsars. This has been noted by \cite{Verbunt2014} who studied the relationships between the types of GCs and the pulsars they host. They classified all GCs with known pulsars based on their interaction rate per single binary and noted that GCs with a high value of this parameter show unusual properties like: a higher percentage of isolated pulsars, binary pulsars that have acquired massive and exotic companions through exchange interactions and pulsars with long periods or unusually short estimated characteristic ages. The suggested reason behind this is that the high interaction rate per binary leads to an increased disruption of binaries and/or an exchange of the companions.
In the list provided by \cite{Verbunt2014}, NGC 6522 appears as the one with the highest interaction rate per single binary ($\sim 4$ times that of Terzan 5 and $\sim 10$ times that of 47 Tucanae). This would explain the observed population of pulsars.

These arguments have been put forth as motivating factors in further observing this GC. The possibility of finding exotic MPSs with unusual companions like neutron stars or even possibly black holes have pushed astronomers to further observe NGC 6522 in recent years \citep{Ridolfi2021}. While \cite{Zhang2020} using the Ultra Wideband Low (UWL) receivers at the Parkes telescope was unable to detect new pulsars, \cite{Ridolfi2021} using only the central 44 antennas of the more sensitive MeerKAT radio telescope found the most recent isolated pulsar, PSR J1803$-$3002D. 

In the current work we performed further searches for new pulsars in NGC 6522 in observations made by the large survey projects (LSPs) MeerTIME\footnote{\url{http://www.meertime.org}} \citep{Bailes2016,Bailes2020} and TRansients And PUlsars with MeerKAT  (TRAPUM; \citealt{Stappers2016}) \footnote{\url{http://www.trapum.org}} using the full array of MeerKAT. Thanks to the backends provided by TRAPUM \citep{Barr2018,Chen2021} we could synthesize a large number of coherent beams in the sky with the ability of searching for pulsars in all of them. These capabilities have proven to be extremely useful in observing and localizing new discoveries up to a few arcmin away from the center of a GC \citep{Abbate2022}. This allowed us to cover a region larger than about 2 half-light radii (69 arcsec, \citealt{Baumgardt2018}) without losing sensitivity away from the center.

\section{Observations}

\begin{figure*}
\centering
	\includegraphics[width=0.49\textwidth]{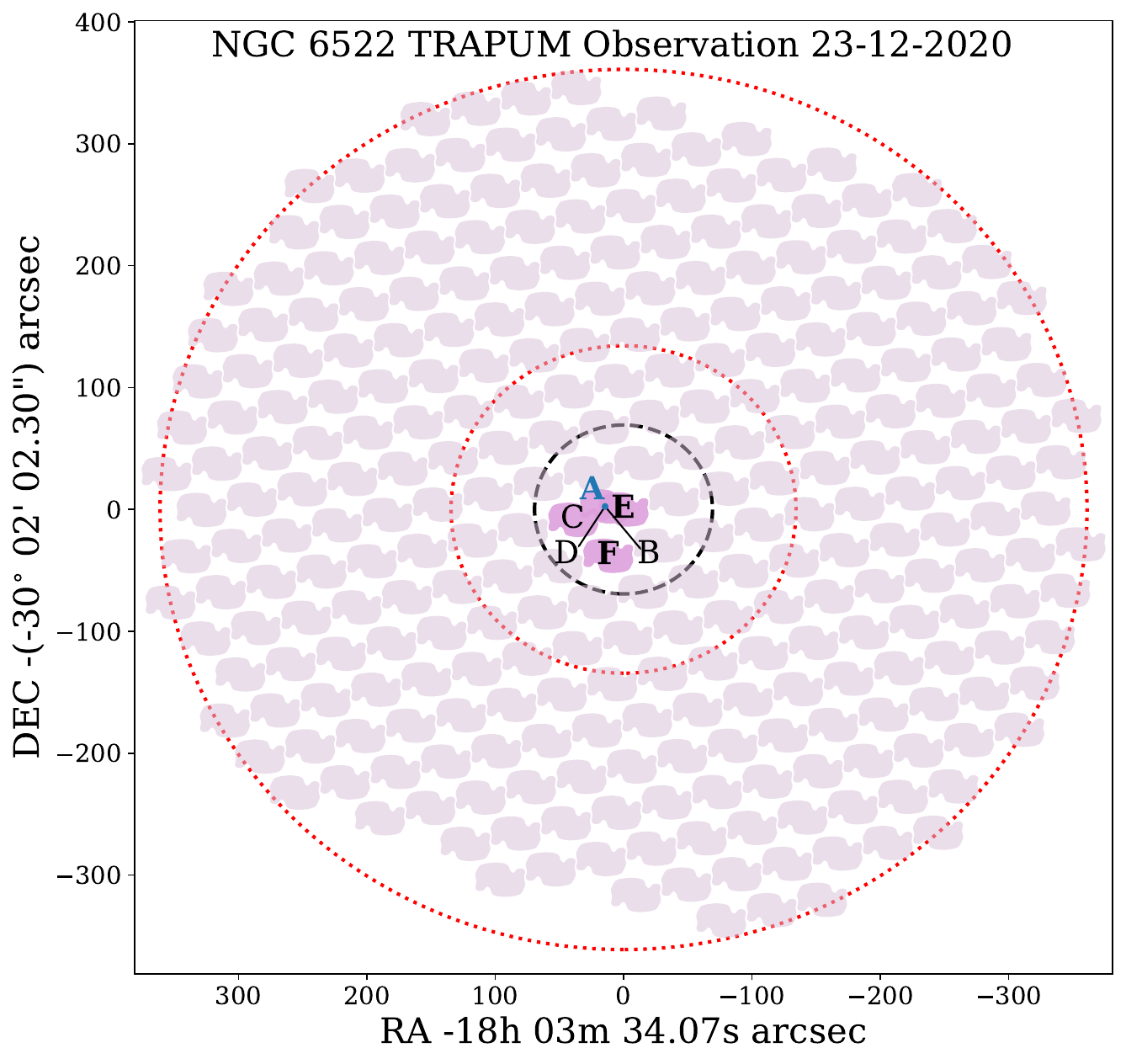}
	\,
	\includegraphics[width=0.49\textwidth]{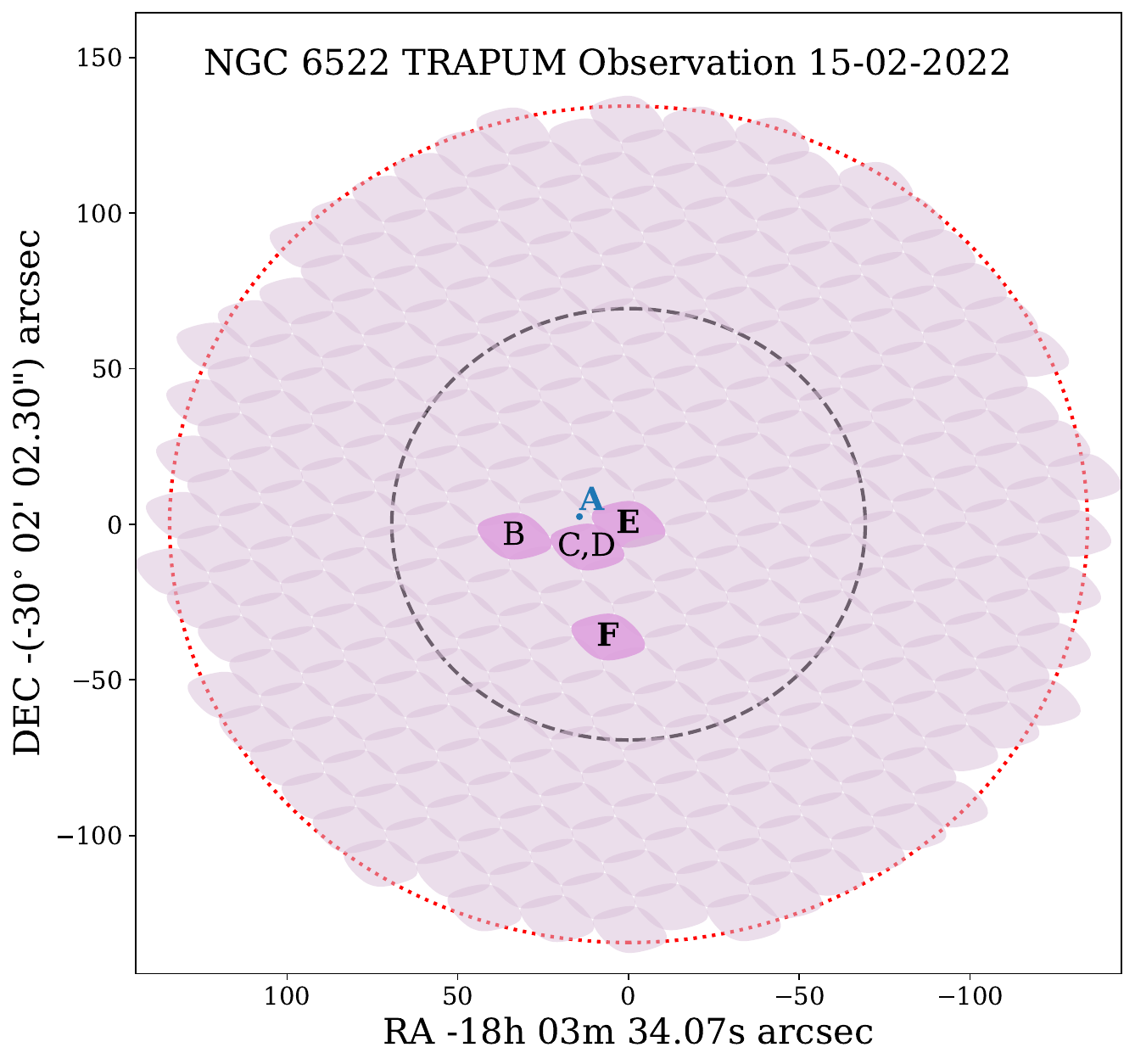}
  	\caption{Tiling pattern of the beams of the TRAPUM observations of NGC 6522 performed on 2020 December 23 (left) and on 2022 February 2022 (right). The tied array beams are shown in pink up to the half-power level measured at 1284 MHz at the middle of the observation. The black dashed circle shows the half-light radius of the cluster, $\sim 1.15$ arcmin \citep{Baumgardt2018}. The red dotted circles show the maximum extent of the size of the tiling in the two observations plotted for comparison. PSR J1803$-$3002A is shown in blue at the position measured through timing \citep{Zhang2020} while the other pulsars are shown at position of the beam where they appeared with the highest S/N. The beams with the best detections are represented in a brighter shade of pink. The new discoveries are shown in bold. All of the beams shown have been searched for new pulsars.}
  	\label{fig:tiling_patterns}
\end{figure*}

The observations at the MeerKAT telescope used to search for new pulsars were performed in the L-band (856-1712 MHz) on 2020 December 23 and on 2022 February 15. In both cases the number of antennas used was 60, the number of tied array beams formed in the sky was 288 and the observation length was 4 h. The shape and tiling pattern of the tied array beams for the two observations were determined using the software \texttt{MOSAIC}\footnote{\url{https://github.com/wchenastro/Mosaic}}\citep{Chen2021} and are shown in Fig. \ref{fig:tiling_patterns}. For the first observation, the single tied array beam at 50 percent of the power and calculated at the middle point of the observation at a frequency of 1284 MHz (the central frequency of the observing band) can be approximated by an ellipse with semimajor axis of $\sim 20$ arcsec and semiminor axis of $\sim 15$ arcsec. The 288 beams covered a region up to $\sim 6$ arcmin in radius. For the second observation the single tied array beam at 50 percent of the power can be approximated by an ellipse with semimajor axis of $11$ arcsec and a semiminor axis of $\sim 7.5$ arcsec. The beams were spread in a way to cover $\sim 2.2$ arcmin. Despite using the same number of antennas, the beams of the first observation are almost twice as large as the second observation and are distributed to cover an area almost 9 times as large. This can be explained by the fact that in the first observations the antennas with the largest baselines were excluded for technical reasons while they were available for the second observation. Because of the different areas covered by the observations, we can only confirm candidates discovered in the region covered by both observations. Nonetheless, we performed the search analysis on all the beams in order to acquire a list of potential pulsar candidates to look for in upcoming observations.

We made use of the Filterbanking Beamformer User Supplied Equipment (FBFUSE) and Accelerated Pulsar Search User Supplied Equipment (APSUSE) backends provided by TRAPUM \citep{Barr2018}. For each coherent beam we recorded the total intensity using the filterbank format with a time resolution of 76 $\mu$s and a total bandwidth of 856 MHz centered around 1284 MHz with 4096 frequency channels. After the recording was finished, the data were incoherently dedispersed at the value of DM corresponding to PSR J1803-3002A, 192.6 \pccm \citep{Zhang2020}. The data were successively cleaned from bright radio frequency interference (RFI) using the Inter-Quartile Range Mitigation algorithm\footnote{\url{https://github.com/v- morello/iqrm}} \citep{Morello2021}. After these steps the frequency channels were averaged by a factor of 16 down to 256 for the following analysis and long-term storage.

\section{Search analysis}

The search analysis was carried out using the \texttt{PRESTO} \citep{Ransom2002} based \texttt{PULSAR\_MINER}\footnote{\url{https://github.com/alex88ridolfi/PULSAR_MINER}} pipeline described in \cite{Ridolfi2021}. The first step of the processing was an additional search for RFI still present in the data through the \texttt{rfifind} routine and a search for periodic RFI. These RFI are found through a Fourier transform of the time series created at 0-DM and are later removed from the Fourier spectra during the actual search. 

The DM range where pulsars were previously found in the cluster is 192-194 pc cm$^{-3}$ \citep{Possenti2005,Begin_2006,Ridolfi2021}. In order to not miss any pulsar we searched the range 184-200 pc cm$^{-3}$ using step sizes of 0.05 \pccm.
Given the value of de-dispersion chosen for the observations, 192.6 \pccm, and the reduced number of channels, pulsars with a DM at the edge of the searched range will be affected by smearing. For the worst affected channel, the smearing caused by the largest offset of DM would be $\sim 0.3$ ms. This implies that our sensitivity to rapidly spinning can be affected only if the DM is significantly offset from the central value.

The search is performed for isolated and for binary pulsars in the Fourier domain for spin periods in the range 1 ms to 20 s. In case of binary pulsars, the orbital acceleration causes the power spectrum to be spread along several Fourier bins determined by the value $z=t_{\rm obs}^2 a_l/(cP)$, where $t_{obs}$ is the length of the segment of observation searched, $a_l$ is the line-of-sight acceleration due to the orbital motion and $P$ is the spin period of the pulsar. When searching for binary pulsars, we considered signals spread up to a number of Fourier bins of $z_{\rm max}=200$.
This assumption is valid only as long as the acceleration along the line of sight stays constant but it breaks down if the observation covers a fraction of the orbit longer than around 10 percent \citep{Ransom2003}.
In order to be sensitive to shorter orbits we divided the observation in smaller segments of 20 and 60 min and repeated the search in each segment. Therefore, for the accelerated search, we are sensitive to orbits longer than $\sim 200$ min. 

To improve the sensitivity to more compact binaries we also employed the technique of `jerk search' \citep{Eatough2013,Bagchi2013,Andersen2018}. This algorithm looks for linear changes in acceleration (written as $\dot a$) in the Fourier spectrum by a number of Fourier bins determined by $w= \dot a T_{\rm obs}^3 /(cP)$. We extended the search up to $w_{\rm max}= 600$. Because of the significantly higher computational requirements of the `jerk search' we only applied it to the central beam of the observation for both epochs. Since the population of pulsars is usually concentrated towards the center of the cluster, this is the beam where the probability of finding new pulsars is the highest.

Out of all of the candidates found with these methods, we kept only the ones with a Fourier significance higher than 4$\sigma$ (for details see \citealt{Ransom2002}) for visual inspection. In order to confirm a candidate, we checked if the same candidates showed up in the neighbouring beams or in beams with compatible sky position from the other observation. For interesting candidates we tried to fold neighbouring beams and the beams of the other observation at the interesting values of period and DM. 

After a pulsar has been confirmed, we extracted an approximate ephemeris and folded the observations using the \texttt{DSPSR}\footnote{\url{http://dspsr.sourceforge.net}} package \citep{vanStraten2011}. Next, we improved on the value of DM of the pulsar. After extracting frequency-dependent times-of-arrival (ToAs) integrated in time using the \texttt{PAT} routine of \texttt{PSRCHIVE}\footnote{\url{http://psrchive.sourceforge.net}} \citep{Hotan2004,vanStraten2012}, we fitted for the DM using \texttt{TEMPO2}\footnote{\url{https://bitbucket.org/psrsoft/tempo2/}} \citep{Hobbs2006}. Finally, we tried to localize the pulsars using the information from the multiple beam detections using \texttt{SEEKAT}\footnote{\url{https://github.com/BezuidenhoutMC/SeeKAT}} \citep{Bezuidenhout2023}. This code performs a maximum likelihood estimation to obtain the best pulsar position by comparing the telescope point spread function measured using \texttt{MOSAIC} with the S/N of the detections in each beam.

%Furthermore, we look for these pulsars in archival observations of the cluster with the intent of obtaining a phase-connected timing solution.

%flux calibration
 Finally, the flux density of the pulsars was measured using the radiometer equation \citep{Handbook2012}:

\begin{equation}
    S= {\rm (S/N)_{\rm best}} \frac{T_{\rm sys}}{G\sqrt{n_p t_{\rm obs} \Delta f}}\sqrt{\frac{W}{P-W}},
\end{equation}
where (S/N)$_{\rm best}$ is the signal to noise ratio (S/N) of the best detection of the pulsar, $T_{\rm sys}$ is the system temperature of the telescope taken to be 26~K for the observations\footnote{\url{https://skaafrica.atlassian.net/rest/servicedesk/knowledgebase/latest/articles/view/277315585}}, $G=(N_{\rm ant}/64)\times 2.8$ K~Jy$^{-1}$ is the gain of the telescope and depends on $N_{\rm ant}$ the number of antennas used in the observation, $n_p=2$ is the number of polarizations summed, $t_{\rm obs}$ is the length of the observation, $\Delta f$ is the effective bandwidth taken to be 700 MHz after the RFI cleaning and $P$ is the period of the pulsar. $W$ is the equivalent width of the pulse defined as the width of a top-hat function with the same height and same total area as the pulsar profile.

\section{Results}

\subsection{Discoveries}

\begin{table*}
\caption{Properties of the newly discovered pulsars in NGC 6522. 
}
\label{tab:discoveries}
\footnotesize
\centering
\renewcommand{\arraystretch}{1.0}
\vskip 0.1cm
\begin{tabular}{lccccllc}
%\hline
%\multicolumn{7}{c}{Summary of Discoveries}\\
\hline

Pulsar name     & Type        & \multicolumn{1}{c}{Reference epoch} & \multicolumn{1}{c}{$P$}   &  \multicolumn{1}{c}{DM}        &  \multicolumn{1}{c}{$\alpha_{\rm J2000}$}  & \multicolumn{1}{c}{$\delta_{\rm J2000}$} & \multicolumn{1}{c}{S$_{1283}$}  \\
\vspace{0.3cm}
       &             & \multicolumn{1}{c}{(MJD)}& \multicolumn{1}{c}{(ms)}   & \multicolumn{1}{c}{(pc cm$^{-3}$)}  &  &  & \multicolumn{1}{c}{($\mu$Jy)}\\
\hline

J1803$-$3002E   &  Isolated   &  59205 & 17.926  & 192.79(4)  & 18$\h$03$\m$34$\fs$14$^{+0.3}_{-0.2}$ & $-30\degr02\arcmin07\arcsec$$^{+11}_{-2}$ & 5.0 \\ 
J1803$-$3002F   &  Isolated   & 59625 & 148.136   & 195.8(5)  & 18$\h$03$\m$34$\fs$7$^{+0.3}_{-0.2}$ & $-30\degr02\arcmin35\arcsec$$^{+2}_{-7}$ & 6.7 \\ 

\hline
\end{tabular}
\vspace*{2mm}
\caption*{{\bf Notes:} We report the barycentric period, DM, position and flux at 1283 MHz. The numbers in parentheses represent 1-$\sigma$ uncertainties in the last digit. For the position we report the location of the highest peak in the \texttt{SEEKAT} localization and the 1-$\sigma$ interval.}
\end{table*}

\begin{figure*}
\centering
	\includegraphics[width=0.4\textwidth]{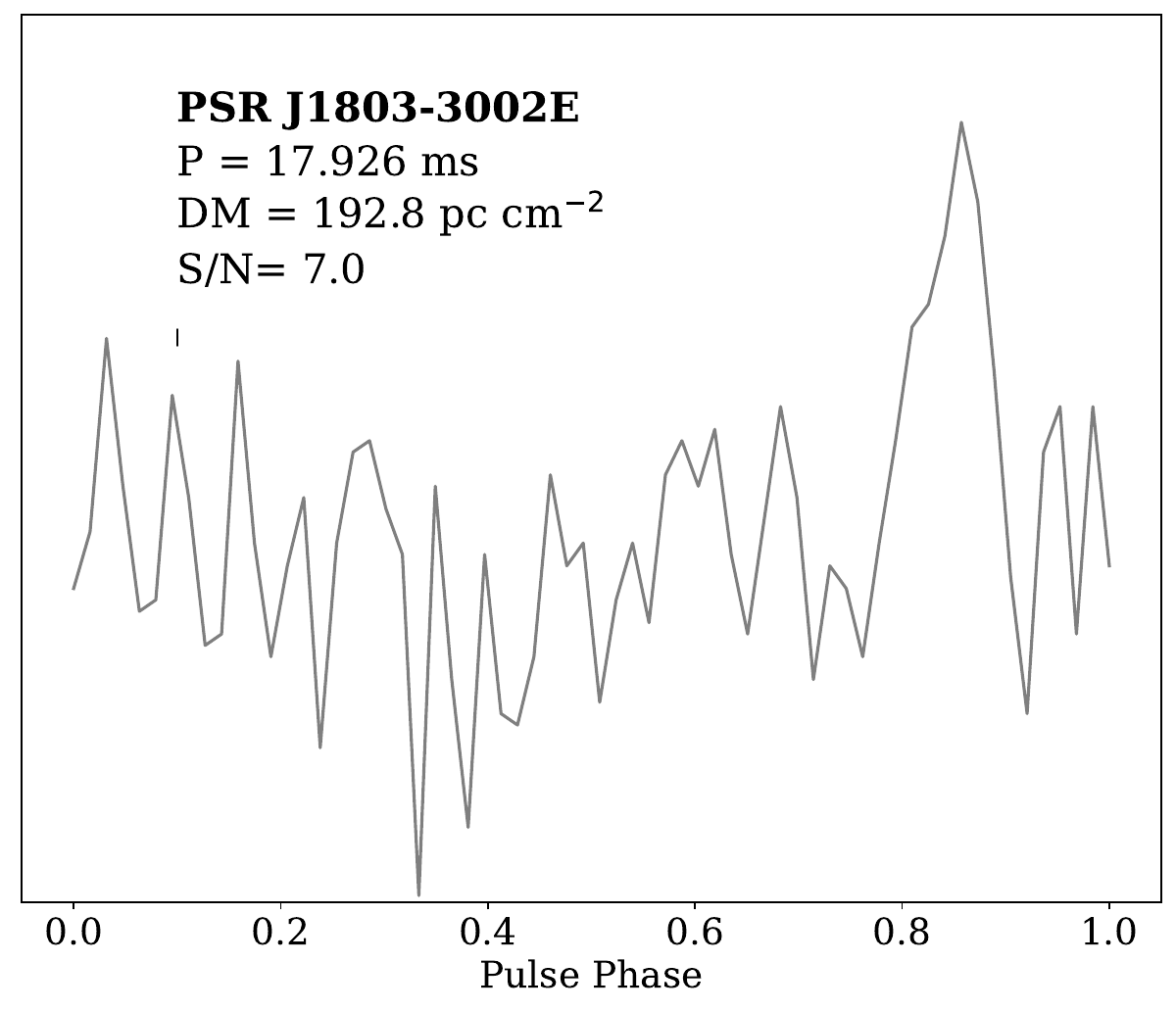}
	\,
	\includegraphics[width=0.4\textwidth]{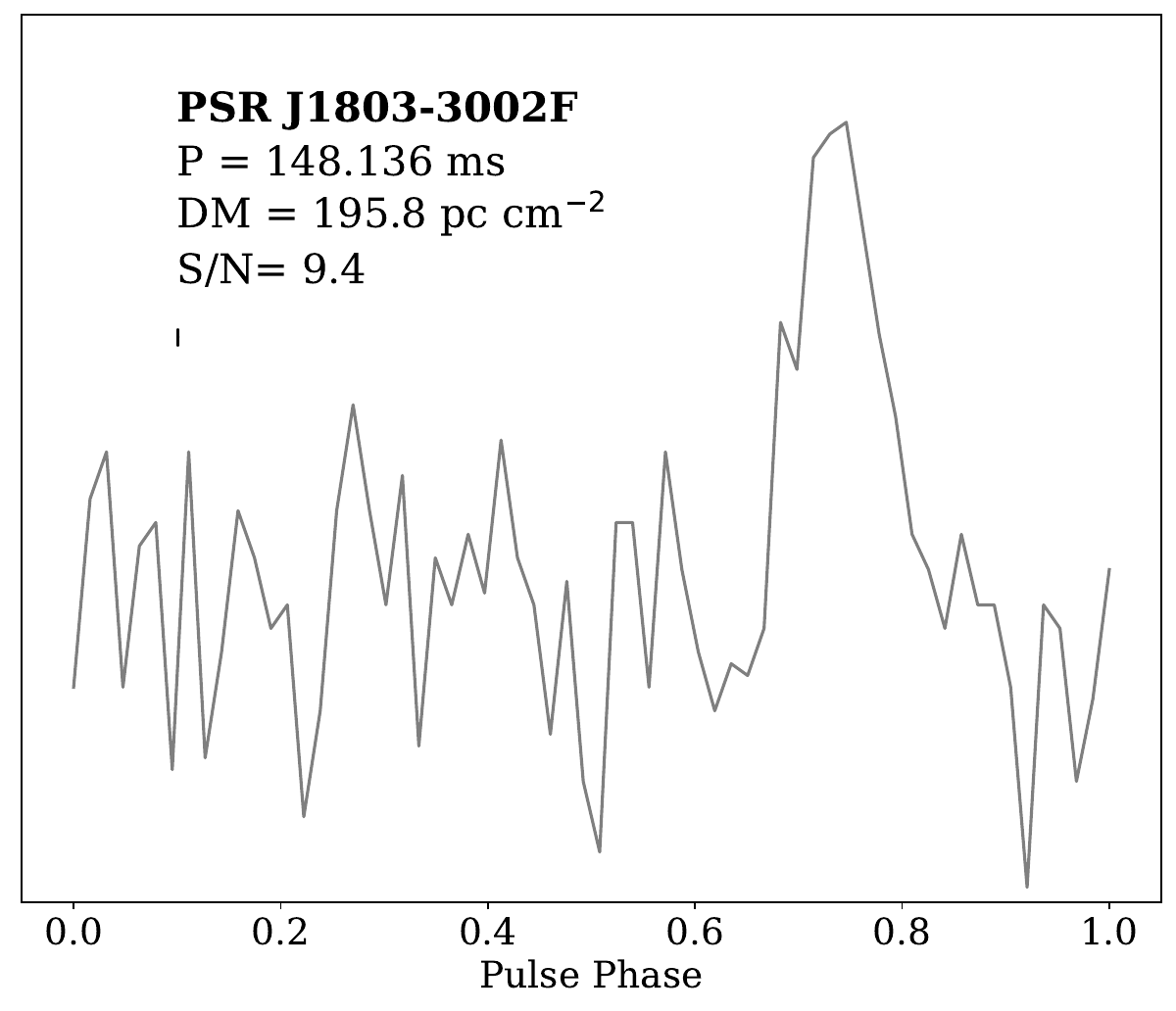}
	\,
  	\caption{Profiles of the two pulsars discovered in the TRAPUM observations of NGC 6522 at L-band. The name, period, DM and S/N for each pulsar are reported in each plot. The bar under the DM shows the quadrature sum of the time resolution of 76 $\mu$s of the observations and the DM smearing in the worst affected channel in comparison to the period of the pulsars. The profiles are obtained from folding the the beam with the highest S/N in the discovery observation. In both cases the integration time is 4 h.}
  	\label{fig:discovery_profiles}
\end{figure*}

The search analysis was run on all of the beams for both isolated and accelerated pulsars while the `jerk search' was run only on the central beams of both observations. Two pulsars were discovered: PSR J1803$-$3002E and PSR J1803$-$3002F. The properties and profiles of the two new pulsars are shown in Table \ref{tab:discoveries} and in Fig. \ref{fig:discovery_profiles}. The localization of the new discoveries with \texttt{SEEKAT} is shown in Fig. \ref{fig:seekat_positions}.

PSR J1803$-$3002E is a 17.9 ms pulsar with a DM of 192.8 pc cm$^{-3}$ found originally in the central beam of the first observation. The profile has a single peak with equivalent width of 1.7 ms. It was later redetected in another beam of the same observation, in 7 beams of the observation that occurred on 2022 February 15 and in an older MeerKAT observation of the cluster of 2019 October 16 described in \cite{Ridolfi2021}. It does not show any acceleration during the observations and the value of the period does not change significantly over the three observations meaning that the pulsar is likely isolated. From the errors of the spin periods measured for the different epochs using the \texttt{PDMP} routine of \texttt{PSRCHIVE}, we can provide an upper limit on the period derivative of $\sim 3 \times 10^{-14}$ ss$^{-1}$. The localization map of the pulsar shown in Fig. \ref{fig:seekat_positions} is non-symmetric and shows two possible peaks. This is caused by the small number of beams where the pulsar was detected and low S/N of each detection. Nonetheless the location of the pulsar close to the center of the cluster and the DM value close to the other pulsars in the cluster confirm that the pulsar is a member of the cluster. 

PSR J1803$-$3002F is a 148 ms pulsar with a DM of $195.8\pm 0.5$ pc cm$^{-3}$ found in the second observation and then redetected in 6 neighbouring beams and in one beam of the first observation corresponding to the same position. The profile has a single peak with equivalent width of 13.9 ms. It also appears to be isolated as it does not show significant acceleration during the observations and the spin period at both epochs is consistent. Also in this case the errors on the localization shown in Fig. \ref{fig:seekat_positions} are asymmetric due to similar reasons as for PSR J1803$-$3002E.
The pulsar is found within the half-light radius of the cluster and the DM difference with the rest of the pulsars ($\sim 3$ \pccm) is consistent with the observed DM spread seen in other GCs\footnote{The properties of the known pulsars in all GCs can be found at the webpage: \url{https://www3.mpifr-bonn.mpg.de/staff/pfreire/GCpsr.html}}. Therefore we can conclude that also this pulsar is a member of the cluster.

\subsection{Detection of previously known pulsars}

\begin{figure*}
\centering
	\includegraphics[width=0.8\textwidth]{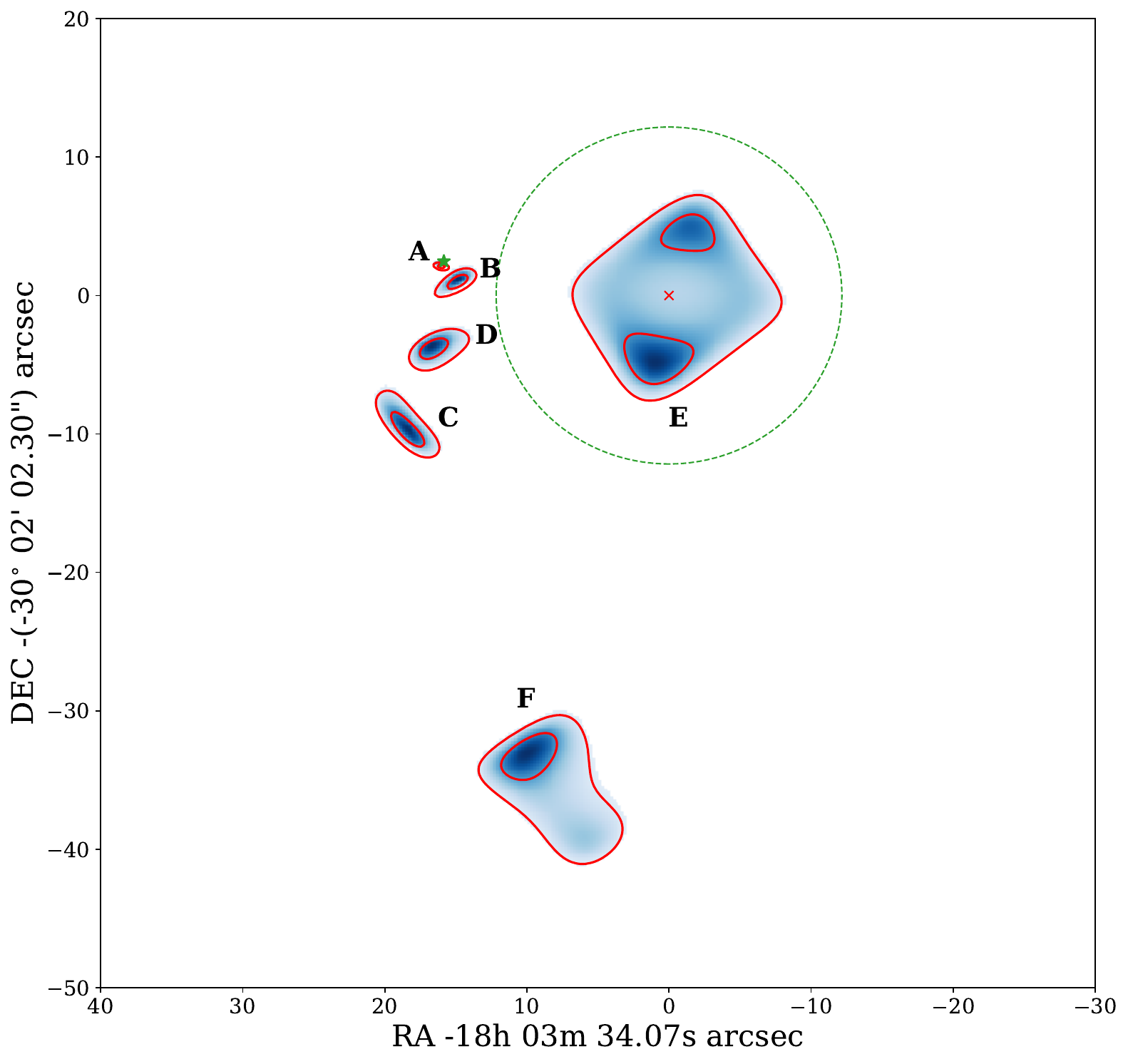}
  	\caption{Localization of known pulsars and the new discoveries in NGC6522 using \texttt{SEEKAT}. We show the likelihood map of the \texttt{SEEKAT} localization showing the 1$\sigma$ and 2-$\sigma$ confidence intervals. The larger and asymmetric uncertainties for PSR J1803$-$3002E and PSR J1803$-$3002F are caused by the low S/N of the detections and the low number of beams used for the localization. The position of PSR J1803$-$3002A determined by timing \citep{Zhang2020} is shown as a green star and is compatible with the \texttt{SEEKAT} localization of the pulsar. The red cross shows the location of the optical center of the GC, while the green dashed circle shows the core radius of the cluster of $\sim 12.3$ arcsec \citep{Kerber2018}. }
  	\label{fig:seekat_positions}
\end{figure*}

We used the same observations and the previous ones performed with the MeerKAT telescope and described in \cite{Ridolfi2021} to look at the other pulsars of the cluster. While PSR J1803$-$3002A has been studied thoroughly by \cite{Zhang2020} using the Parkes telescope, and therefore has a timing position, the same work was not able to detect the other pulsars thus their positions are unknown. In our observations with MeerKAT we were able to redetect all of the previously known pulsars in multiple beams. This allowed us to attempt a localization of them using \texttt{SEEKAT} and the results are shown in Fig. \ref{fig:seekat_positions} and in Table \ref{tab:known_pulsars}. In the same table we also report a more accurate value of DM measured from our detections. Given that the previous MeerTIME observations described in \citep{Ridolfi2021} also retained Stokes values, we also attempted a calculation of the rotation measure (RM). We tried to use both the \texttt{RMFIT} routine of \texttt{PSRCHIVE} and the code described in \cite{Abbate2023}. However, we were not able to measure an RM for any of them suggesting that the linear polarization fraction of these pulsars is low.

\begin{table*}
\caption{Properties of the previously known pulsars in NGC 6522.
}
\label{tab:known_pulsars}
\footnotesize
\centering
\renewcommand{\arraystretch}{1.0}
\vskip 0.1cm
\begin{tabular}{lcllc}

\hline
Pulsar name     &  \multicolumn{1}{c}{$\alpha_{\rm J2000}$}  & \multicolumn{1}{c}{$\delta_{\rm J2000}$} &\multicolumn{1}{c}{DM} & \multicolumn{1}{c}{S$_{1283}$}\\
\vspace{0.3cm}
       & &  & \multicolumn{1}{c}{(pc cm$^{-3}$)} & \multicolumn{1}{c}{($\mu$Jy)}\\
\hline
J1803$-$3002B   &   18$\h$03$\m$35$\fs$06(5) & $-30\degr02\arcmin01\farcs$4(4)  & 191.79(2) & 22 \\ 
J1803$-$3002C   &   18$\h$03$\m$35$\fs$3(2) & $-30\degr02\arcmin12\arcsec$(1)  & 193.88(5) & 18 \\ 
J1803$-$3002D   &   18$\h$03$\m$35$\fs$06(5) & $-30\degr02\arcmin01\farcs$4(4)  & 192.66(2) & 13 \\ 
\hline
\end{tabular}
\vspace*{2mm}
\caption*{{\bf Notes:} We report the position, the new measure of DM {and the flux at 1283 MHz}. The numbers in parentheses represent 1-$\sigma$ uncertainties in the last digit. For the position we report the location of the highest peak in the \texttt{SEEKAT} localization and the 1-$\sigma$ interval.}
\end{table*}

\subsection{Timing of PSR J1803$-$3002C}

\begin{figure}
\centering
	\includegraphics[width=0.45\textwidth]{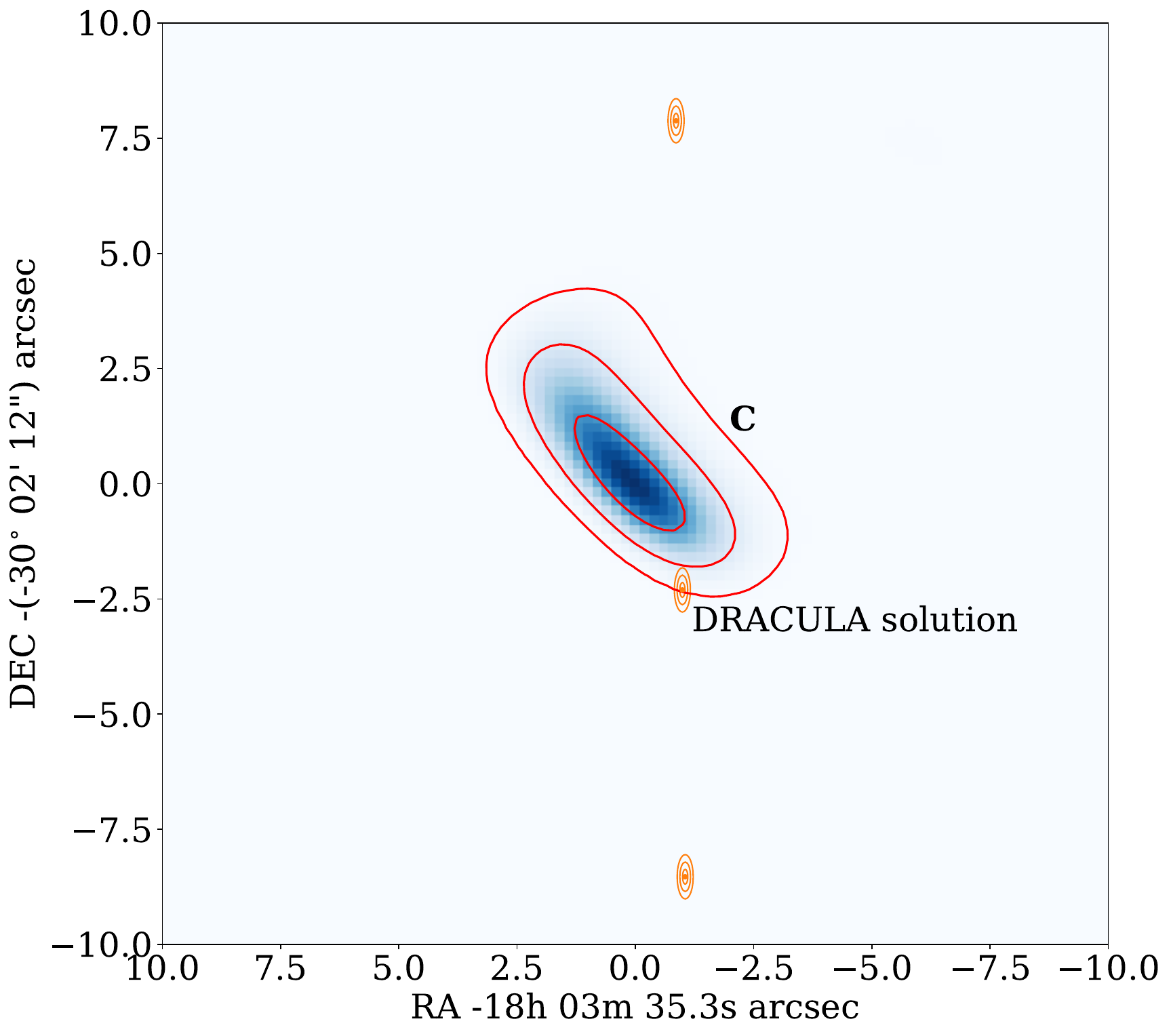}
	\,
  	\caption{Position of the solution found by \texttt{DRACULA} that is closest to the \texttt{SEEKAT} localization of PSR J1803$-$3002C. The red contours represent the 1$\sigma$, 2$\sigma$ and 3$\sigma$ levels of the \texttt{SEEKAT} localization. The orange dot represents the position of the closest timing solution found with the 1$\sigma$, 2$\sigma$  and 3$\sigma$ levels shown with orange contours. The two positions overlap at the 3$\sigma$ levels. On the North and South side we show also the two next closest timing solutions.}
  	\label{fig:timing_C_position}
\end{figure}

\begin{figure}
\centering
	\includegraphics[width=0.48\textwidth]{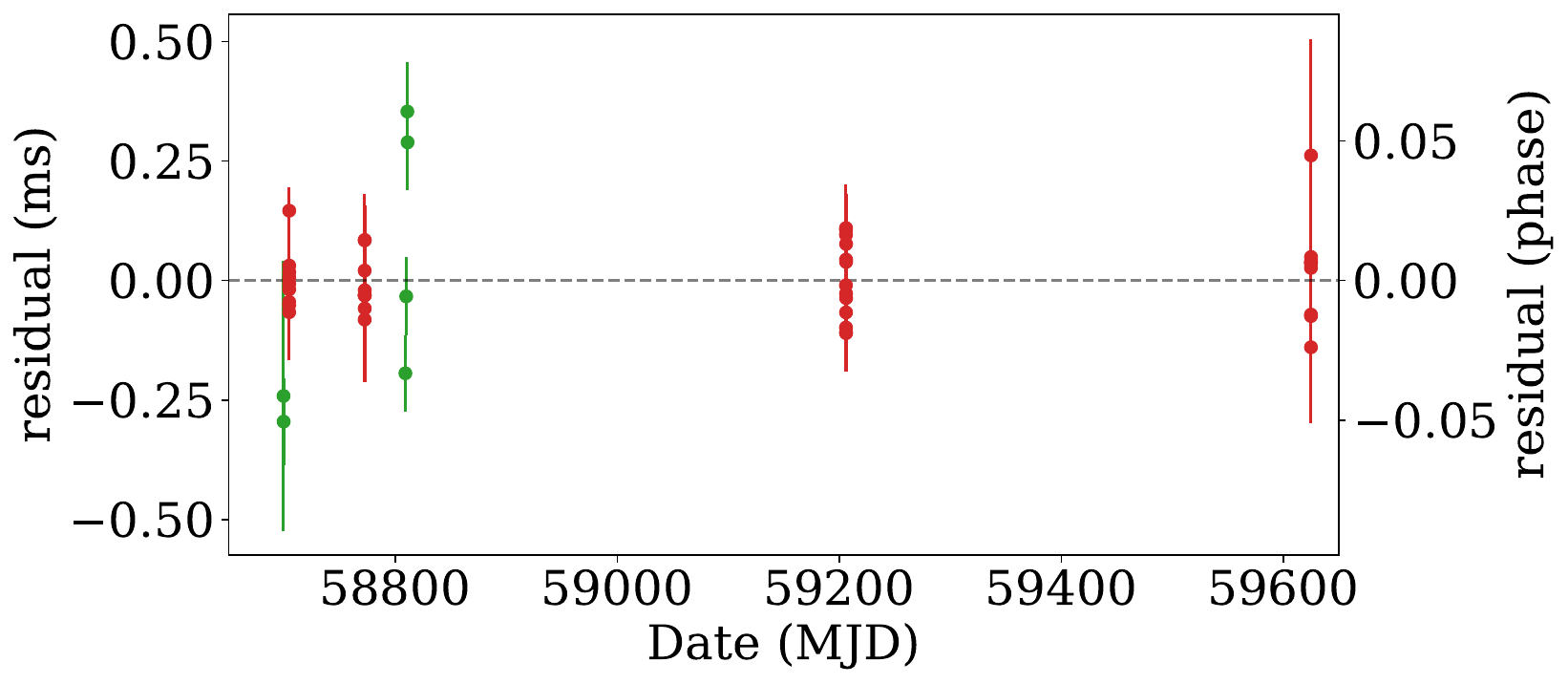}
	\,
  	\caption{ToA residuals of the timing solution found by \texttt{DRACULA} closest to the localization found by \texttt{SEEKAT}. The red points show the ToAs extracted from MeerKAT observations while the green points show the ToAs extracted from Parkes UWL observations.}
  	\label{fig:residuals_C}
\end{figure}

Spurred by the redetection of these pulsars, we tried to fold the pulsars in the observations taken using the Ultra Wideband Low (UWL) receiver at the Parkes telescope, described in \cite{Zhang2020}. In the four observations taken in search mode we were able to detect PSR J1803$-$3002C in all of the observations and PSR J1803$-$3002D in only one. From these detections we were able to extract ToAs to complement the ones taken from the observations with MeerKAT. We tried to obtain a phase connected timing solution using the \texttt{DRACULA}\footnote{\url{https://github.com/pfreire163/Dracula}} \citep{Freire2018} code based on \texttt{TEMPO}\footnote{\url{https://tempo.sourceforge.net}} that is especially useful for sparse data sets. Running the code on the ToAs extracted for PSR J1803$-$3002C while fitting only for the position, spin period and spin period derivative, returned 743 different possible solutions
with a reduced $\chi^2$ smaller than 2. We checked the position in the sky of these timing solutions and found that only one was close enough to the position determined by \texttt{SEEKAT} to be compatible at the 3$\sigma$ level. This solution is shown in Fig. \ref{fig:timing_C_position} while the residuals of the ToAs of this solution are shown in Fig. \ref{fig:residuals_C}. The parameters of the timing solution are shown in Table \ref{tab:timing_fitresults}. The next two closest solutions are both at a distance of $\sim 8$ arcsec from the center of the \texttt{SEEKAT} localization and are also shown in Fig. \ref{fig:timing_C_position} for reference.

The characteristic age shown in the table and calculated using the spin period and spin period derivative is surprisingly small, $\tau_c = 133$ Myr. This is in contrast with the old age of the stars in the cluster \citep{Barbuy2009,Kerber2018}. The other two closest timing solutions found by \texttt{DRACULA} shown in Figure \ref{fig:timing_C_position} have similar values of characteristic ages of 126 and 139 Myr. We checked whether the cluster's gravitational potential might have a significant contribution to the spin period derivative. The cluster radial density profile can be described accurately with a King profile \citep{Kerber2018} so we can use the equation for the acceleration derived in \cite{Freire2005} to find the maximum possible value at the position of the pulsar. The core radius derived from the King profile is $12.3 \pm 0.3$ arcsec \citep{Kerber2018} while the stellar velocity dispersion is $\sim 8.2$ km s$^{-1}$ \citep{Baumgardt2018}. Using a distance of 7.3 kpc \citep{Baumgardt2021}, we get a maximum value of the cluster-induced spin frequency derivative of $\sim 2.5\times 10^{-15}$ Hz s$^{-1}$, almost eight times smaller than the measured value. Furthermore, the position at $\sim 1.5$ core radii from the center makes it unlikely that a non-luminous mass in the center of the GC could influence so drastically the acceleration. Alternatively, the presence of a nearby star whether in a wide binary or passing by could influence the period derivative. A longer timing baseline is needed to test these hypotheses. The most likely explanation could be that the spin-down is intrinsic. Similar young characteristic ages have been measured for other pulsars in GCs, mostly slow or mildly recycled pulsars \citep{Verbunt2014} with the exception of the MSPs PSR J1823$-$3021A and PSR B1821$-$24A. However, both of these MSPs are very bright and are visible also in the gamma-ray band, while PSR J1803$-$3002C is relatively weak and does not appear in the latest catalog of Fermi Large Area Telescope sources \citep{Abdollahi2022}.

\begin{table*}
\setlength\tabcolsep{10pt}
    \caption{Timing solution for PSR J1803$-$3002C found by \texttt{DRACULA}.}
    \centering
     \label{tab:timing_fitresults}
    \renewcommand{\arraystretch}{1.0}
     \begin{tabular}{l c} 
     \hline
      \hline
Pulsar  &   J1803$-$3002C    \\
\hline
R.A. (J2000) \dotfill   &  18$\h$03$\m$35$\s$.221(4)    \\
DEC. (J2000) \dotfill   & $-30\degr02\arcmin14\arcsec.5$(1)   \\
Spin Frequency, $f$ (s$^{-1}$)  \dotfill 	& 171.22104938954(5) \\
1st Spin Frequency derivative, $\dot{f}$ (Hz\,s$^{-1}$) \dotfill & $-$2.044(1)$\times 10^{-14}$\\
Reference Epoch (MJD)  \dotfill & 59206.206359\\
Start of Timing Data (MJD) \dotfill & 58699.375 \\
End of Timing Data (MJD) \dotfill 	& 59625.230 \\
Dispersion Measure (pc\,cm$^{-3}$) \dotfill	& 193.88\\
Number of ToAs \dotfill     & 59\\
Weighted rms residual ($\mu$s) \dotfill  &  103.325 \\
%$S_{1300}$ (mJy) \dotfill &  0.039\\
\hline
\multicolumn{2}{c}{Derived Parameters}  \\
\hline
Spin Period, $P$ (s)   \dotfill & 0.00584040340(2)\\ 
1st Spin Period derivative, $\dot{P}$ (s\,s$^{-1}$)  \dotfill &  6.973(4)$\times 10^{-19}$ \\
Surface Magnetic Field, $B_0$, (G)  \dotfill & 2.04 $\times 10^{9}$ \\
Characteristic Age, $\tau_{\rm c}$ (Myr) \dotfill & 133 \\
    \hline
      \hline
     \end{tabular}
\vspace*{2mm}
\caption*{{\bf Notes:} The timing solution for is derived from the timing of the observed ToA with \texttt{DRACULA} closest to the position of the pulsar found by \texttt{SEEKAT}. The time units are TDB, the adopted terrestrial time standard is TT(TAI) and the Solar System ephemeris used is JPL DE421 \citep{Folkner2009}. Numbers in parentheses represent 1-$\sigma$ uncertainties in the last digit.}
\end{table*}

\section{Discussion}\label{sec:disc}

The spin period of both newly discovered pulsars is higher than the other pulsars in the cluster: PSR J1803$-$3002E could be classified as a mildly recycled pulsar and PSR J1803$-$3002F has a long spin period. 
Among all pulsars in the version 1.70 of \texttt{psrcat}\citep{Manchester2005} catalogue\footnote{\url{https://www.atnf.csiro.au/research/pulsar/psrcat/}} with spin period within 0.1 and 0.2 s, the median value of the characteristic age is $\sim 5 \times 10^5$ yr and the 95 percent confidence interval is ($5 \times 10^3$- $4 \times 10^9$) yr. This suggests that the characteristic age of PSR J1803$-$3002F is likely to be smaller than the age of the cluster. Furthermore, the timing of PSR J1803$-$3002C suggests that this is also a relatively young pulsar. These types of pulsars are usually found only in GCs with high interaction per single binary as shown by \cite{Verbunt2014}. In the same work, the authors showed that these GCs have a higher fraction of isolated pulsars with respect to binaries. Finding two slow isolated pulsars in addition to the previously known four isolated MSPs in NGC 6522, classified as having the highest interaction rate per single binary in the sample studied by \cite{Verbunt2014}, supports this thesis. 

The presence of these slow and young pulsars in GCs has a number of proposed explanations. The analysis of \cite{Verbunt2014} suggests that these pulsars were formed when a stellar interaction disrupted the X-ray binary where this NS was located. This premature halting of the recycling process can happen before the magnetic fields of the pulsars were ablated to the levels seen in most MSPs. In this interpretation, the apparently young characteristic age of the pulsars corresponds roughly to to the epoch when the accretion was halted. The high occurrence rate of these disruptive interactions in core-collapsed GCs could be the cause of the lack of binary pulsars and predominance of isolated ones. 

An alternative solution is the recent formation of young pulsars following the accretion induced collapse of massive white dwarfs \citep{Tauris2013,Kremer2023}. This scenario would be more efficient in core-collapsed clusters where simulations show a large number of white dwarfs might migrate towards the center enhancing the chance of collisions or mergers \citep{Kremer2021}. In this scenario, the slow and young pulsars should be more concentrated in the center of the GCs where the bulk of the white dwarfs are located given that the natal kicks should be of order 10 km s$^{-1}$ \citep{Kremer2023b}. The position of PSR J1803-3002F slightly offset from the center but still within the half-light radius is compatible with this interpretation.

In order to test these scenarios we would need an estimate of the spin period derivatives of these pulsars from which we can estimate the surface magnetic field and the characteristic age. In the scenario of disrupted recycling, the pulsar should be located below the adjusted spin-up line in the P-${\rm \dot P}$ diagram of \cite{Verbunt2014}, like all other apparently young pulsars in GCs \citep{Abbate2022}. If a pulsar is over that line than it must have been born recently. In the case of PSR J1803$-$3002C the estimated period derivative places it below the spin-up line leaving both possibilities open. For the case of PSR J1803$-$3002E and PSR J1803$-$3002F more observations of the cluster are needed to obtain the spin period derivative through a timing solution.

\section{Conclusions}

In this manuscript we report the discovery of two new isolated pulsars in NGC 6522 in observations made with the MeerKAT telescope. PSR J1803$-$3002E is a mildly recycled pulsar with a spin period of 17.926 ms found near the center of the GC while PSR J1803$-$3002F is a slow pulsar with a spin period of 148.136 ms at $\sim 3$ core radii from the center. With these new discoveries, the total number of known pulsars in this cluster is six, all of which are isolated. The unusual absence of any binary pulsar in the cluster can be attributed to the high disruption rate of binaries in the cluster. Indeed, NGC 6522 is one of the GCs with the highest interaction rate per single binary and these types of GCs show a larger percentage of isolated pulsars \citep{Verbunt2014}.

Using the same observations it was possible to redetect and provide a more accurate value of the DM for the other known pulsars in the cluster. Thanks to the multi-beam detection of the TRAPUM observations, we were also able to localize the pulsars.

Combining all of the observations at MeerKAT of NGC 6522 with the ones made at the Parkes telescope, it was possible to derive a tentative timing solution for PSR J1803$-$3002C. While the timing solution found by \texttt{DRACULA} is not unique, only one solution is found to be compatible with the localization of the pulsar performed by \texttt{SEEKAT}. This solution appears to have a very high spin period derivative which implies a small characteristic age of only $\sim 132$ Myr. After considering the expected contribution caused by the GC's gravitational potential, the spin period derivative is likely to be intrinsic to the pulsar. If confirmed, the pulsar would have one of the smallest characteristic age among the MSPs associated with GCs. 

While not yet certain, the apparent young age of PSR J1803$-$3002C would be particularly interesting in light of the new discoveries presented here. The presence of both a young MSP and slow pulsars in the same cluster, similarly to the case of NGC 6624, would suggest that the formation of these unexpected objects is linked as described in Section \ref{sec:disc}. This is in line with the predictions of \cite{Verbunt2014}. An alternative idea behind the origin of these pulsars is that they were born recently after the collapse of white dwarfs \citep{Tauris2013,Kremer2023}.
A larger sample of these objects in more GCs is needed to prove their origin.

Only with future observations of the cluster can we confirm the tentative timing solution of PSR J1803$-$3002C, find timing solutions for the other pulsars in the cluster and discover new potentially interesting pulsars. 
Counter-intuitively, the prevalence of isolated pulsars should motivate us to look more intensively for highly accelerated binary pulsars. Clusters with mostly isolated pulsars are sometimes found to host binaries that result from exchange encounters like the case of PSR J1823$-$3021G \citep{Ridolfi2021} in NGC6624 and PSR B2127+11C \citep{Anderson1990} in M15. Among these systems it is possible to find exotic binaries like double neutron star systems. Thanks to their high eccentricities, these systems are ideal for measuring masses and potentially test general relativity using post-Keplerian parameters.
These systems should be the main targets for future surveys of NGC 6522.

\begin{acknowledgements}

The MeerKAT telescope is operated by the South African Radio Astronomy Observatory, which is a facility of the National Research Foundation, an agency of the Department of Science and Innovation. SARAO acknowledges the ongoing advice and calibration of GPS systems by the National Metrology Institute of South Africa (NMISA) and the time space reference systems department department of the Paris Observatory. TRAPUM observations used the FBFUSE and APSUSE computing clusters for data acquisition, storage and analysis. These clusters were funded and installed by the Max-Planck-Institut für Radioastronomie and the Max-Planck-Gesellschaft.
PTUSE was developed with support from the Australian SKA Office and Swinburne University of Technology. MeerTime data is housed on the OzSTAR supercomputer at Swinburne University of Technology. The OzSTAR program receives funding in part from the Astronomy National Collaborative Research Infrastructure Strategy (NCRIS) allocation provided by the Australian Government. The authors also acknowledge MPIfR funding to contribute to MeerTime infrastructure.
The Parkes radio telescope is part of the Australia Telescope National Facility (https://ror.org/05qajvd42) which is funded by the Australian Government for operation as a National Facility managed by CSIRO. We acknowledge the Wiradjuri people as the Traditional Owners of the Observatory site.
FA, AR, PCCF, PVP, VB, MK, EDB, WC, DC acknowledge continuing valuable support from the Max-Planck Society. This work is supported by the Max-Planck Society as part of the ``LEGACY'' collaboration on low-frequency gravitational wave astronomy. 
AR and AP gratefully acknowledge financial support by the research grant ``iPeska'' (P.I. Andrea Possenti) funded under the INAF national call Prin-SKA/CTA approved with the Presidential Decree 70/2016. Part of this work has been funded using resources from the INAF Large Grant 2022 ``GCjewels'' (P.I. Andrea Possenti) approved with the Presidential Decree 30/2022. LZ is supported by ACAMAR Postdoctoral Fellowship and the National Natural Science Foundation of China (Grant No. 12103069). The National Radio Astronomy Observatory is a facility of the National Science Foundation operated under cooperative agreement by Associated Universities, Inc. SMR is a CIFAR Fellow and is supported by the NSF Physics Frontiers Center award 2020265.

\end{acknowledgements}
% WARNING
%-------------------------------------------------------------------
% Please note that we have included the references to the file aa.dem in
% order to compile it, but we ask you to:
%
% - use BibTeX with the regular commands:
\bibliographystyle{aa} % style aa.bst
\bibliography{biblio} % your references Yourfile.bib
%
% - join the .bib files when you upload your source files

%-------------------------------------------------------------------
\end{document}